\documentclass[aps,preprint,showpacs]{revtex4}
\usepackage{graphicx}
\usepackage{dcolumn}

\newcommand{\tauiso}{{\mbox{\boldmath $\tau$}}}
\newcommand{\bm}{\bibitem}

\begin{document}

\title{Dilepton production in nucleon-nucleon collisions reexamined}
\date{\today}
\author{R. Shyam$^{1,2}$ and U. Mosel$^3$}
\affiliation {
$^1$ Saha Institute of Nuclear Physics, Kolkata 700064, India \\
$^2$ Theory Center, Thomas Jefferson National Accelerator Facility, 
12000 Jefferson Avenue, Newport News, VA 23606, USA\\ 
$^3$ Institut f\"ur Theoretische Physik, Universit\"at Giessen, D-35392 
Giessen, Germany } 

\date{\today}

\begin{abstract}
We present a fully relativistic and gauge invariant framework for 
calculating the cross sections of dilepton production in nucleon-nucleon 
($NN$) collisions which is based on the meson-exchange approximation for 
the $NN$ scattering amplitudes. Predictions of our model are compared with 
those of  other covariant models that have been used earlier to describe 
this reaction. Our results are also compared with those of the semiclassical
models which are employed to get the input elementary cross sections in the
transport model calculations of the dilepton production in nucleus-nucleus 
collisions. It is found that cross sections obtained within the semiclassical
and quantum mechanical models differ noticeably from each other. 
\end{abstract}
 
\maketitle
\newpage
\section{Introduction}

A recurring feature of the dilepton ($e^+e^-$) spectra measured in 
nucleus-nucleus ($AA$) collisions has been the enhancement (above known 
sources) in the invariant mass distribution of the cross sections in the 
region of the vector meson ($\rho^0$ and $\omega$ ) pole mass. This has been
the case for experiments performed for bombarding energies ranging from as 
low as 1 GeV/nucleon (DLS data~\cite{por97}), through the SPS energies 
(40 - 158 GeV/nucleon)~\cite{aga05,ada03,arn06} to the energies employed 
by the PHENIX collaboration at RHIC (which correspond to invariant mass of 
200 GeV/nucleon)~\cite{afa07}. The enhancement seen at the SPS energies are 
understood in terms of the modification of the $\rho$ meson spectral function
in the hadronic medium \cite{rap07}.

However, the large dileptons yields observed in the DLS experiment (in 
$^{12}$C + $^{12}$C and $^{40}$Ca + $^{40}$Ca collisions at 1-2 GeV/nucleon
beam energies) in the invariant mass ($M$) range from 0.2 GeV to 0.5 GeV,  
are yet to be explained satisfactorily~\cite{ern98,cas99,rap00,she03,aga07}. 
Independent transport model calculations have been unable to describe these 
data fully even after including contributions from (i) the decay of $\rho$ 
and $\omega$ mesons which are produced directly from the nucleon-nucleon 
($NN$) and pion-nucleon scattering in the early reaction phase~\cite{bra98},
(ii) the in-medium $\rho$ spectral functions~\cite{cas98}, (iii) the 
dropping $\rho$ mass with corresponding modification in the resonance 
properties~\cite{ern98}, (iv) an alternative scenario of the in-medium 
effect - a possible decoherence between the intermediate meson states in 
the vector resonance decay \cite{she03}. This led to term this discrepancy 
as "DLS-puzzle" \cite{ern98,bra98} which persists even now. 

In order to resolve this unsatisfactory situation the high-acceptance 
dielectron spectrometer (HADES) has been built~\cite{aga07} which allows to
study the dilepton production in elementary proton-proton ($pp$), 
proton-deuteron ($pd$) as well as in proton-nucleus ($pA$) and $AA$ collisions
with much wider acceptance region for beam energies up to 8 GeV/nucleon. 
Unlike the DLS experiment HADES also measures the dilepton yields in the 
quasi-free proton-neutron ($pn$) scattering. The first set of data has already
been published~\cite{aga07,aga08} by this group on $^{12}$C + $^{12}$C 
collisions at beam energies of 1.0 and 2.0 GeV/nucleon. The remarkable 
fact is that these data agree well with those of the DLS collaboration. 
Therefore, there is no longer any question against the validity of the DLS 
data and the previous failures to explain them by various transport models 
have to do with problems in the theoretical calculations.

On the theory side, in a recent HSD transport model calculation~\cite{bra08}
it has been shown that if one uses larger cross sections for elementary $pp$
and $pn$ bremsstrahlung processes the observed dilepton yields in the 
relevant invariant mass region for the $^{12}$C + $^{12}$C collisions at 
1-2 GeV/nucleon can be reproduced. The support for the enhanced elementary 
bremsstrahlung cross sections comes from the calculations of these processes
presented in Ref.~\cite{kap06} within a model which is similar to that used 
in Ref.~\cite{shy03}. Althought the calculations perfomed within the two 
models use the same input parameters yet the cross sections of ~\cite{kap06} 
are larger than those of Ref.~\cite{shy03} by factors of 2-4. It has been 
further argued in Ref.~\cite{bra08} that elementary bremsstrahlung cross 
sections larger than those of Ref.~\cite{shy03} were reported already in Ref.
\cite{dej96} within a similar type of model which employes however, realistic 
$T$ matrices to describe the initial nucleon-nucleon ($NN$) collisions. 

In this paper, we examine the issue of the dilepton production in 
the elementary $NN$ collisions in order to highlight and understand the 
differences seen in the predictions of various models
\cite{dej96,shy03,kap06} for the corresponding cross sections. This is 
important for a proper theoretical description of the new data on the 
dilepton production in elementary $pp$ and $pn$ collisions which are likely
to be announced soon by the HADES collaboration. This will also have vital 
implications for the predictions of the different transport models for the 
HADES dilepton yields in $AA$ collisions. Indeed, in a very recent UrQMD 
transport model analysis of the HADES $C$ + $C$ data~\cite{san08}, it has 
been shown that even without using the enhanced elementary bremsstrahlung 
cross sections of Ref.~\cite{kap06} the observed dilepton yields can be 
described fairly well at the 2.0 GeV/nucleon beam energy in the region of 
relevant $M$ values.  Although, this theory still underpredicts the data in
this mass region at the 1.0 GeV/nucleon beam energy.  
 
The major difference between models of Refs.~\cite{dej96,shy03,kap06} 
lies in the method of implementing the gauge invariance of the $NN$ 
bremsstrahlung amplitudes. To investigate this issue, we have recalculated
the cross sections for the dilepton production in elementary $pp$ and $pn$ 
reactions within a fully relativistic and gauge invariant model which is 
similar to that used in Refs.~\cite{shy03,kap06} [to be referred as full 
quantum mechanical model (FQM)] except for the fact that we have used a 
pseudoscalar (PS) nucleon-nucleon-pion ($NN\pi$) vertex instead of the
pseudovector (PV) one employed by these authors. The reason is that with 
the PS coupling, the contact term (seagull diagram) is not involved in the
total Lagrangian which however, still remains gauge invariant for bare 
point like nucleons. With a PV $NN\pi$ coupling one needs to introduce a 
contact term to restore the gauge invariance of the total amplitude and 
different approaches~\cite{pen02} for constructing this contribution lead
to very different results~\cite{uso05}. The use of the PS $NN\pi$ vertex 
makes our discussions free from this ambiguity. 

In the computation of the amplitudes, strong form factors are introduced
to quench the contributions from high momenta and to include effects due to
the compositeness of the nucleon. However, this leads to the loss of gauge 
invariance. We show here that form factors at various vertices (hadronic as 
well as electromagnetic) can still be implemented in our model without loosing
gauge invariance. We make a detailed comparison of the ingredients and 
predictions of FQM with those of the models of Refs.~\cite{dej96,kap06} and 
examine their validity  critically. The predictions of the FQM  is also 
compared with those of the semiclassical models which are used to obtain the 
elementary $NN$ bremsstrahlung and delta isobar contributions  to the 
elementary dilepton production reactions in transport models to calculate 
the cross sections for this reaction in $AA$ collisions. 
 
\section{Formalism}
 
\begin{figure*}
\begin{center}
\includegraphics[width=0.4 \textwidth]{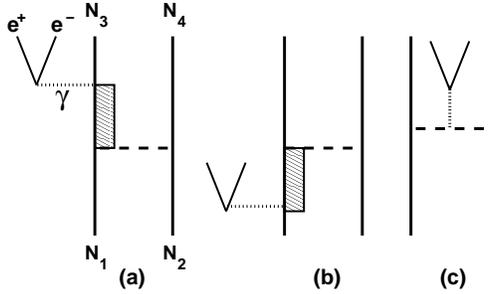}
\end{center}
\vskip -.2in
\caption{
A representative of Feynman diagrams for emission of dilepton
in nucleon-nucleon collision as considered in this work.
(a) denotes emission after $NN$ collisions, (b) before $NN$ collision
and (c) during $NN$ collision. The box represents an off-shell
nucleon or a $\Delta$ isobar. 
}
\end{figure*}

A representative of the lowest order Feynman diagrams contributing to
the dilepton production within our model is shown in Fig.~1. The 
intermediate nucleon or resonances can radiate a virtual photon which decays 
into a dilepton (Figs.~1a and 1b). There are also their exchange counterparts.
In addition, there are diagrams of similar types where the virtual photon is 
emitted from the nucleon line on the right side.  The internal meson lines 
can also lead to dilepton emission (see Fig 1c). To evaluate various 
amplitudes, we have used the same effective Lagrangians for the 
nucleon-nucleon-meson, resonance-nucleon-meson, nucleon-nucleon-photon and 
resonance-nucleon-photon vertices as discussed in Ref.~\cite{shy03} with the
sole exception that for the $NN\pi$ vertex we have used the PS 
coupling
\begin{eqnarray}
{\cal L}_{NN\pi} & = & ig_{NN\pi} {\bar{\Psi}}_N \gamma _5
                             \tauiso \cdot {\bf \Phi}_\pi \Psi _N, 
\end{eqnarray} 
instead of the PV one.  The parameters of the model were taken to 
be the same as those of Ref.~\cite{shy03} where details of 
calculations of various amplitudes are given.  The on-shell equivalence 
of the pseudoscalar and pseudovector couplings makes the parameters
independent of the choice for the type of the $NN\pi$ vertex.

In the computation of the amplitudes corresponding to diagrams shown in 
Fig.~1 the problem of loosing the gauge invariance
arises while using the electromagnetic form factors if the complete vertex 
function for the half-off-shell photon production is not used
\cite{sch94,tie90,don95}. The full vertex is given by
\begin{eqnarray}
\Gamma^\mu (p^\prime,p) & = & e \sum_{i=1}^{3} \sum_{s,s^\prime=\pm 1}
     \Lambda^{s^\prime} F_i^{s^\prime,s}(W^\prime,W;q^2){\cal O}_i^{\mu}
 \Lambda^s(p)
\end{eqnarray} 
where operators ${\cal O}$ are defined as ${\cal O}_1^\mu = \gamma^\mu$,
${\cal O}_2^\mu = -i\sigma^{\mu\nu}q_\nu /2m_N$ and ${\cal O}_3^\mu = -q^\mu$
with $m_N$ being the nucleon mass. $\Lambda$'s are projection operators
which are given by
\begin{eqnarray}
\Lambda^s(p)& = &\frac {s\gamma \cdot p +W}{2W}, 
\end{eqnarray}
with $W = \sqrt{p^2}$ and $s = \pm 1$. In Eq. (2) $F_i$ are the form factors.
In these equations $p$ and $p^\prime$ denote the four momenta of the incoming
and outgoing nucleons, respectively and $q$ is the four momentum of the photon.
We can relate $F_1$ and $F_3$ by using Ward-Takahashi identity (WTI)
~\cite{sch94}
\begin{eqnarray}
q_\mu \Gamma^\mu (p^\prime,p) = e\frac{(1+\tau_3)}{2}[S^{-1}(p^\prime) - 
                                S^{-1}(p)],
\end{eqnarray}
where $S(p)$ is the nucleon propagator. It should be noted that WTI does not 
pose any constraint on the magnetic form factor $F_2$. 

It is easy to see that Eq.~(4) will restore gauge invariance for the $pn$ 
bremsstrahlung for the exchange of an uncharged meson (see, e.g. 
Ref.~\cite{sch94}). The current conservation condition (CCC) 
$q_\mu \Gamma^\mu$ = 0 is satisfied if the contributions of diagrams 1(a) and 
1(b) are added together. In case of the exchange of charged mesons however, 
the sum of these two diagrams does not vanish and one has to also include 
the contributions from the diagram 1(c) to fulfill the CCC.

In actual calculations the hadronic vertices also contain strong form
factors that depend on the four momenta of the exchanged mesons. Therefore,
for uncharged mesons, it is sufficient to multiply all the vertices
by the same form factor as that of the hadronic vertices in order to keep
the gauge invariance. On the other hand, the four momentum of the meson 
changes in Fig.~1(c) for the case of the charged meson exchange and the 
corresponding meson current no longer satisfies the continuity equation even
after being multiplied by the same hadronic form factor. One needs to 
multiply this form factor in diagram 1(c)~\cite{hag89} by an additional 
factor $F(\Lambda_\phi)$ which is obtained by letting $F(\Lambda_\phi)$ 
multiply the mesonic current and then solving for the continuity equation. 
This leads to~\cite{tow87}
\begin{equation}
F(\Lambda_\phi)= 1 + \frac{m_\phi^2 - q_1^2}{\Lambda_\phi^2 - q_2^2} + 
    \frac{m_\phi^2 - q_2^2}{\Lambda_\phi^2 - q_1^2},
\end{equation}
where $\Lambda_\phi$ is the cut-off parameter and $q_1$ and $q_2$ are the
four momentum transfers at the left and right vertices of graph 1(c), 
respectively. This result can be interpreted as the photon coupling to the 
regular pion (first term) and to "heavy" pion at the left and right vertices
(second and third terms, respectively). This way of gauging the strong form 
factor makes it possible to use a given form factor for the meson and a 
different one for the nucleon but still fulfill the WTI.

As far as form factors for the electromagnetic vertices of the nucleons are
concerned,  we note that for real photons, the gauge invariance mandates
$F_1$ = 1, $F_2$ = $\kappa_N$ and $F_3$ = 0, where $\kappa_N$ is the 
magnetic moment of the nucleon. For the actual case there is a considerable
uncertainty in these form factors. $F_3$ is never accessible by experiments 
since ${\cal O}_3^\mu j_\mu$ = 0 for any conserved current.  As in previous 
studies~\cite{sch94,shy03,kap06,feu99} we have chosen not to include 
electromagnetic form factors for the nucleon. 

We now compare our model with those of Refs.~\cite{dej96,kap06}. In 
Ref.~\cite{dej96}, instead of the one-boson exchange picture of our model, 
the nucleon-nucleon interaction is included via a $T$-matrix that is based 
on the Paris potential. However, the nucleon current is not gauge invariant 
in this model. These authors rectify this problem in an ad-hoc manner which 
may not have a microscopic basis. In the model of Ref.~\cite{kap06}, a 
pseudovector $NN\pi$ coupling has been used. With this coupling gauge 
invariance is preserved with a contact term  ($NN\pi\gamma$ vertex) added to 
the total Lagrangian. However, once hadronic form factors are introduced the 
gauge invariance will be violated and a gauge restoration procedure has to be 
applied~\cite{uso05}. 

In Ref.~\cite{kap06} the same method as described above, has been used 
for restoring gauge invariance in the case of $pp$ collisions. No contact 
graph due to pseudovector coupling appears in this case. For the $np$ case 
however, a contact term is needed. Different approaches for constructing the 
additional current contribution (contact term) to restore gauge invariance 
lead to different types of form factors~\cite{pen02} which can yield quite 
different results~\cite{uso05}. The usual practice is to choose a 
prescription which provides best agreement with the data. However, no 
comparison is shown with the experimental data in this paper. 

\section{Results and Discussions}
 
\begin{figure*}
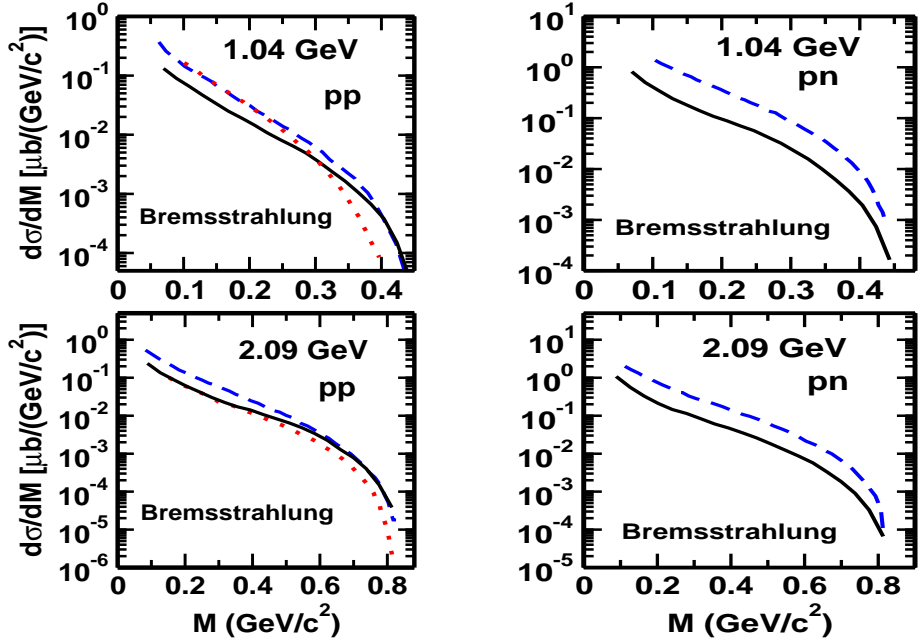

\begin{center}
\begin{tabular}{cc}
\includegraphics[height=8.5cm,width=5.5cm]{Fig2a.eps} & \hspace{0.99cm} 
\includegraphics[height=8.5cm,width=5.4cm]{Fig2b.eps}
\end{tabular}
\end{center}
\vskip -0.2in
\caption{[color online]
The invariant mass distribution of the $NN$ bremsstrahlung contributions to
the dilepton spectra in $pp$ (left) and $pn$ (right) collisions at the beam 
energies of 1.04 GeV and 2.09 GeV. Results obtained within our model are 
shown by solid lines while those of Refs.~\protect\cite{dej96} and 
\protect\cite{kap06} by dotted and dashed lines, respectively.  
}
\end{figure*}
 
In Fig.~2, we show the invariant mass distribution of the $pp$ and $pn$ 
bremsstrahlung contributions to the dilepton spectra at the beam energies of 
1.04 GeV and 2.09 GeV. Also shown here are results for this reaction as 
reported in Refs.~\cite{dej96,kap06}. First of all we remark that the  
cross sections calculated in the present work are very similar to those 
reported in our earlier work~\cite{shy03} - their shapes are unchanged while
absolute magnitudes of the former are slightly larger than those of the 
latter (by about 10$\%$). However,  the cross sections reported for the $pp$ 
case in Ref.~\cite{dej96} (to be referred as dJM) are larger than ours for 
invariant mass ($M$) $<$ 0.25 GeV at the beam energy of 1.04 GeV while they 
are almost identical for 2.09 GeV in this mass region. At both the beam 
energies, dJM results are smaller than ours for $M >$ 0.25 GeV. On the 
other hand, the cross sections of Ref.~\cite{kap06} (to be referred at KK) 
are larger than our results everywhere in the region of $M < 0.6$ GeV. An 
important point to note is that there is no overall multiplicative factor 
that differentiates the results of various models.
\begin{figure*}
\begin{center}
\includegraphics[width=0.48 \textwidth]{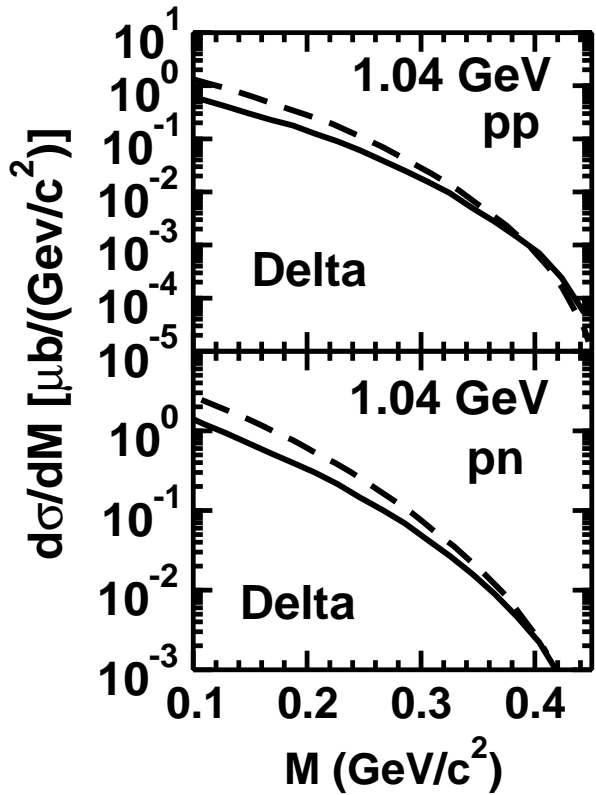}
\end{center}
\vskip -.2in
\caption{[color online]
The invariant mass distribution of the $\Delta$ isobar contribution to the 
dilepton spectra in $pp$ (upper panel) and $pn$ (lower panel) collisions 
the beam energy of 1.04 GeV. The results of our model are shown by full lines 
while those of Ref.~\protect\cite{kap06} by dashed line. 
}
\end{figure*}

Despite using the same diagrams, input parameters and gauge invariance 
restoration procedure, our $pp$ bremsstrahlung cross sections are lower than
those of Ref.~\cite{kap06} as can be seen in the upper left panel of Fig.~2.
Of course, in Ref~\cite{kap06} a pseudovector $NN\pi$ vertex has been used 
as compared to the pseudoscalar one employed in this paper. In this context,
it is worthwhile to note that for the real photon production, the covariant
model calculations do not depend on the choice of the $NN\pi$ coupling 
(PS or PV) as is shown in Ref.~\cite{sch91}. In case of dileptons, different
results can arise for two couplings from the magnetic part of the 
$NN\gamma$ vertex. In fact, in Ref.~\cite{dej97} it is shown that $pp$ 
dilepton bremsstrahlung contributions obtained with the PV $NN\pi$ coupling
are actually smaller than those calculated with the PS one at the beam 
energy of 2 GeV. The calculations presented in Ref.~\cite{shy03} also 
support this to some extent. Because for the $pn$ collisions, dJM results 
are not available, in the right panel of Fig.~2 we compare our results 
with KK cross sections only. We see that latter are larger than those of 
ours by factors ranging between 2-3 at both the beam energies. 

In Fig.~3, we show a comparison of our results and those of Ref~\cite{kap06} 
for the invariant mass distribution of the $\Delta$ isobar contribution to 
the dilepton production in $pp$ and $pn$ collisions at the beam energy of 
1.04 GeV. We note that here too the KK cross sections are larger than ours 
by factors of $\sim$ 2 at smaller values of M even though the two 
models have used the same ingredients and input parameters for this part and 
there is no ambiguity related to gauge invariance as the resonance vertex is 
gauge invariant by its very construction.
\begin{figure*}
\begin{center}
\includegraphics[width=0.48 \textwidth]{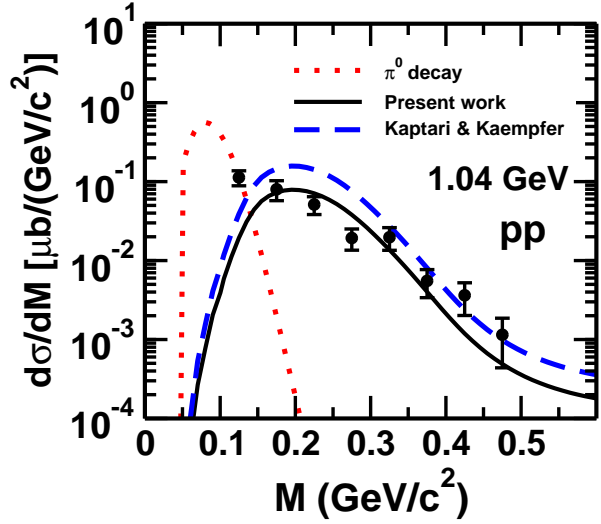}
\end{center}
\vskip -.2in
\caption{[color online]
The calculated dilepton invariant mass distribution for the $pp$ collision at
the beam energy of 1.04 GeV in comparison to the DLS data. The contribution of
the $\pi^0$ Dalitz decay is also shown here which is the same as that in   
Ref.~\protect\cite{shy03}.
}
\end{figure*}

In Fig.~4, we compare the total cross sections of the dilepton production
in $pp$ collisions with the DLS data at the beam energy of 1.04 GeV. The
cross sections calculated within our model are folded with appropriate
experimental filter and final mass resolution. The folded KK cross
sections have been obtained by assuming that the folding procedure does not
affect the ratios of the unfolded cross sections in the two cases. In this
figure we have also shown cross sections for the $\pi^0$ Dalitz decay
($\pi^0 \rightarrow \gamma e^+e^-$) which are the same as those shown in
Ref.~\cite{shy03}. It is seen that KK cross sections overestimate the DLS 
data for $M <$ 0.3 GeV where statistical errors are smaller. The data have 
larger error bars for $M >$ 0.3 GeV. In this context the HADES data on the
elementary dilepton production reactions are expected to be useful because 
of their low statistical error.  

In the several transport model calculations of dilepton production in the 
$AA$ collisions, the usual practice is to calculate the nucleon bremsstrahlung 
contributions within a soft photon approximation (SPA) model~\cite{ruc76,gal87}
and the delta contribution within a Dalitz decay (DDD) model~\cite{wol90}. The
corresponding cross sections are added up to get the total elementary dilepton
production cross sections. In the SPA model, the radiation from the 
internal lines [Fig.~1(c)] is neglected and the strong interaction vertex is 
assumed to be on-shell (which is reasonable only for small photon energies). 
This cross section is corrected  by including a multiplicative factor which 
is the ratio of the phase space available to the two-nucleon system with 
or without the emission of a dilepton of invariant mass $M$~\cite{gal87}. 

It is desirable to check the reliability of the semiclassical calculations  
by comparing them with results of a full quantum mechanical model. In the  
upper panels of Fig.~5, we compare the invariant mass distributions of the 
$pn$ bremsstrahlung cross sections obtained within the SPA and FQM approaches  
at the beam energies of 1.04 GeV and 2.09 GeV. We note that the 
SPA (with corrected phase space) model results agree surprising well in shape 
with those of the FQM. This is however, dependent on the values of the $np$ 
total cross section used in the SPA calculations which is parameterized as 
$\sigma^{np}(s) = \alpha_1m_N/(s-4m_N^2) + \alpha_2$ mb (see, 
Ref.~\cite{gal87}) where $\alpha_1 = 18$ GeV.mb and $\alpha_2 = 10$ mb and
$s$ is the square of the invariant mass in the incident channel. 
\begin{figure*}
\begin{center}
\includegraphics[width=0.48 \textwidth]{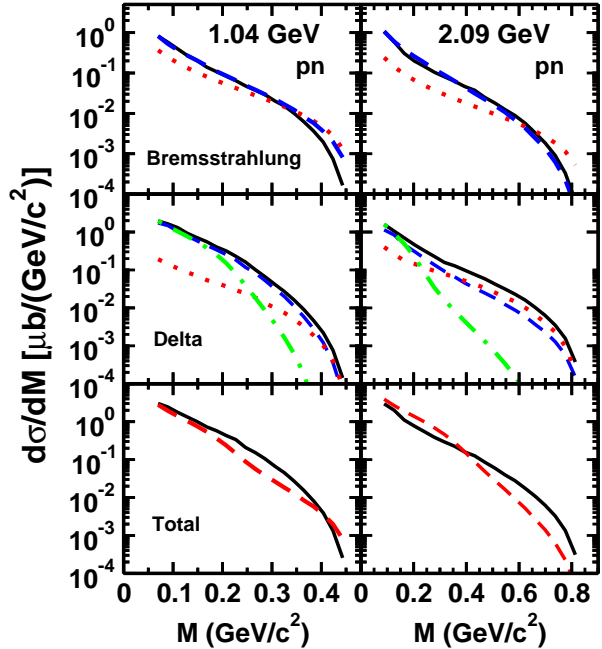}
\end{center}
\vskip -.2in
\caption{[color online] Comparisons of (i) the SPA and FQM $pn$ bremsstrahlung 
contributions [SPA results with and without phase space correction are shown 
by dashed and dotted lines, respectively] (upper panel), (ii) the DDD 
(dashed-dotted lines) and FQM $\Delta$ isobar cross sections (middle panel); 
Also shown here are the FQM post-emission (dashed lines) and pre-emission 
(dotted lines) contributions, (iii) the sum of SPA and DDD contributions 
(dashed line) and the FQM total cross sections (lower panel). The FQM results 
are shown by solid lines.   
}
\end{figure*}

In the middle panel of Fig.~5, we compare the DDD and FQM $\Delta$ isobar 
contributions to the dilepton spectra calculated within two models. We also 
show here the individual contributions of the post-emission (dashed lines) and 
pre-emission (dotted lines) graphs to the FQM cross sections. We note that
at both the beam energies pre-emission diagrams contribute substantially to
the total FQM cross sections for $M > 0.3$ GeV. In the DDD model 
these diagrams are not included. In the lower panel of Fig.~5 we compare the 
total cross sections obtained by adding the SPA and DDD contributions (termed
as semiclassical) with those of the FQM. It is clear that for the important 
intermediate $M$ values the two cross sections differ from each other 
noticeably. Therefore, care has to be taken in interpreting the transport 
model results obtained by using the semiclassical elementary dilepton 
production cross sections.
 
\section{Summary and Conclusions}

In summary, we have presented a fully covariant and gauge invariant model 
for the dilepton production in elementary nucleon-nucleon collisions 
employing a pseudoscalar coupling at the nucleon-nucleon-pion vertex. With 
this coupling, the calculations do not involve the kind of gauge invariance
related ambiguities that are present in those done with a pseudovector
$NN\pi$ coupling. 

We find that similar to the results of Ref.~\cite{shy03}, our $NN$ 
bremsstrahlung cross sections are lower than those of Ref.~\cite{kap06} by 
factors of 2-4 for dilepton invariant mass values below 0.6 GeV for both $pp$ 
and $pn$ collisions at 1.04 GeV as well as 2.09 GeV incident energies. On 
the other hand, the $pp$ bremsstrahlung results of Ref.~\cite{dej96} where 
realistic $T$ matrices have been used to describe the initial $NN$ 
interaction, are similar to those of our model at the beam energy of 2.09
GeV while for 1.04 GeV incident energy they are larger (smaller) than our 
values for dilepton invariant masses smaller (larger) than 0.25 GeV. We 
stress, however, that the current arising from the nucleonic diagram in 
the model of Ref.~\cite{dej96} is not gauge invariant.  An important point 
to note is that there is no overall multiplicative factor to differentiate 
our cross sections from those of Ref.~\cite{dej96}. It is expected that the 
results from the HADES group on the dilepton production in $pp$ and $pd$ 
collisions~\cite{fro07} would provide a fresh ground for differentiating 
between various models. 

Calculations performed with a pseudovector $NN\pi$ coupling are 
more appealing because this coupling is consistent with the chiral symmetry 
requirement of the quantum chromodynamics (QCD)~\cite{wei68} and also because
it leads to negligible contributions from the negative energy states (pair 
suppression phenomena) \cite{mac87}. However, in calculations with PV $NN\pi$
vertex the contact terms resulting from different prescriptions of restoring 
the gauge invariance will have to be carefully examined. This work is  
currently underway.  
  
We found that the total dilepton production cross sections in the 
elementary $NN$ collisions as calculated within the semiclassical models and
used in most transport calculations, differ noticeably from those predicted 
by the full quantum mechanical model. Therefore, quantum mechanical cross 
sections  should be used as input in the transport model calculations of 
dilepton production in nucleus-nucleus collisions in order to interprete 
their results properly.

\section{acknowledgments}

We are grateful to Dr. G. Lykasov and Ingo Fr\"ohlich for a careful reading 
of the manuscript and helpful comments. One of us (RS) would like to thank 
A.W. Thomas for his very kind hospitality at the Theory Center of the Thomas 
Jefferson National Accelerator Facility where a part of this work was done.   
The Jefferson Science Associates (JSA) operates the Thomas Jefferson 
National Accelerator Facility for the United States Department of Energy
under contract DE-AC05-06OR23176.

\end{document}